# SYNCHROTRON OSCILLATION DAMPING DUE TO BEAM-BEAM COLLISIONS*

A. Drago, P. Raimondi, M. Zobov, Istituto Nazionale di Fisica Nucleare, Laboratori Nazionali di Frascati, Via Enrico Fermi 40, Frascati, Italy

Dmitry Shatilov, BINP, Novosibirsk, Russia

*Abstract*

In DAΦNE, the Frascati e+/e- collider, the crab waist collision scheme has been successfully implemented in 2008 and 2009. During the collision operations for Siddharta experiment, an unusual synchrotron damping effect has been observed. Indeed, with the longitudinal feedback switched off, the positron beam becomes unstable with beam currents in the order of 200-300 mA. The longitudinal instability is damped by bringing the positron beam in collision with a high current electron beam (~2A). Besides, we have observed a shift of ≈600Hz in the residual synchrotron sidebands. Precise measurements have been performed by using both a commercial spectrum analyzer and the diagnostics capabilities of the DAΦNE longitudinal bunch-by-bunch feedback. This damping effect has been observed in DAΦNE for the first time during collisions with the crab waist scheme. Our explanation is that beam collisions with a large crossing angle produce a longitudinal tune shift and a longitudinal tune spread, providing Landau damping of synchrotron oscillations.

## INTRODUCTION

In the typical injection scheme of DAΦNE, the electron bunches are stored in the main ring before the positron ones because the electron injection efficiency is higher. Electrons are therefore injected to a rather high current (≈2 A), then the transfer line is switched to positrons, which takes about 1 minute, and then positrons are injected. The electron beam current decays rather rapidly, due to the low energy (510 MeV) and emittance, and finally the two beams collide at approximately the same currents, starting slightly above 1 A.

After electron injection and during the transfer line switching time, there are beam collisions with very high electron currents (between 2A and 1.5A) and relatively low positron ones (between 500 and 200 mA). In this particular situation, a longitudinal damping of the positron beam has been observed even with the longitudinal bunch-by-bunch e+ feedback turned off. This damping effect has been observed in DAΦNE for the first time during collisions with the crab waist scheme [1], [2]. After repeated observations of this behaviour, three dedicated machine studies on Sep/30/09, Nov/04/09 and Nov/05/09, have been carried out with the goal of precisely measuring characteristics of the effect [3].

The instrumentation tools used to make the measurements in the positron main ring have been two: a commercial Real-time Spectrum Analyzer RSA3303A by Tektronix, working from DC to 3 GHz, connected to a high bandwidth beam pickup; and the longitudinal bunch-by-bunch feedback with its beam diagnostic capability both in real time and off-line that can be used in closed and open loop.

## MEASUREMENTS DESCRIPTION

Figure 1 shows a plot from the spectrum analyzer with the 118-th revolution harmonic (highest peak, at 362.484 MHz) of the positron beam together with the synchrotron sidebands placed at 35±1 kHz of distance. The e+ longitudinal feedback is off (i.e. in open loop) and the total beam current is I+ = ~130mA in 103 bunches.

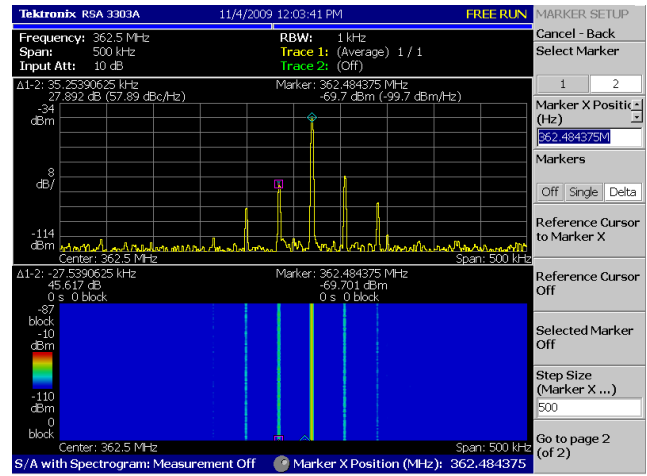

Figure 1: Positron 118-th RF harmonic with synchrotron sidebands.

In the following plot (Fig. 2) the electron beam with a total current of ~1700 mA and all its feedbacks on (i.e. in closed loop) is in collision with e+ beam.

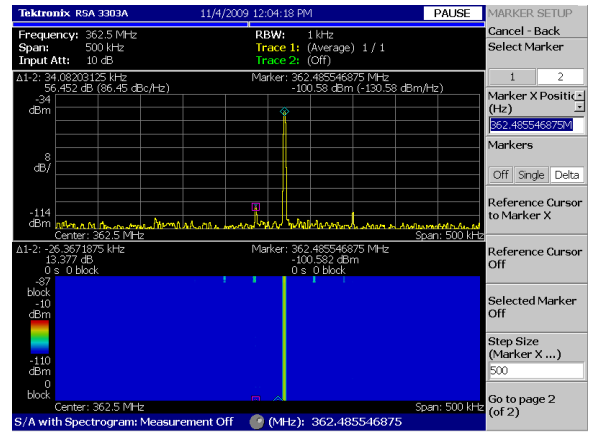

Figure 2: Synchrotron sidebands damped by beam-beam collisions.

The result is clearly shown in the plot: the e+ synchrotron sidebands are almost completely damped, even though the e+ longitudinal feedback is off.

Figure 3 shows the case "in collision – out of collision" with the same setup as the previous ones, that is with the e+ longitudinal feedback turned off and all the others on. A frequency shift of the order of <-1 kHz in the sidebands is clearly visible but the resolution of the instrument is not accurate enough to be exactly measured. It is evident that the damping effect induced by the beam-beam collisions makes lower the synchrotron frequency on both sidebands. In the case of Fig. 3, the total currents are 1550 mA for the electrons and 390 mA for the positrons.

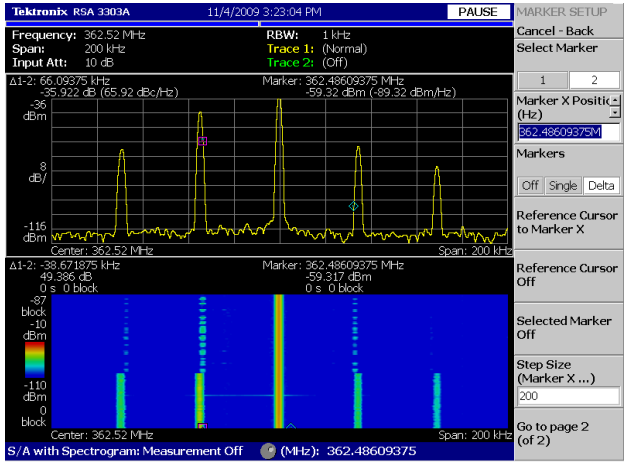

Figure 3: Positron beam longitudinal behaviour shows "in collision – out of collision" case.

It is also possible to download from RSA 3303A the traces also under text form. Downloading these data to the MATLAB environment and zooming the plot, the following Fig. 4 have been created.

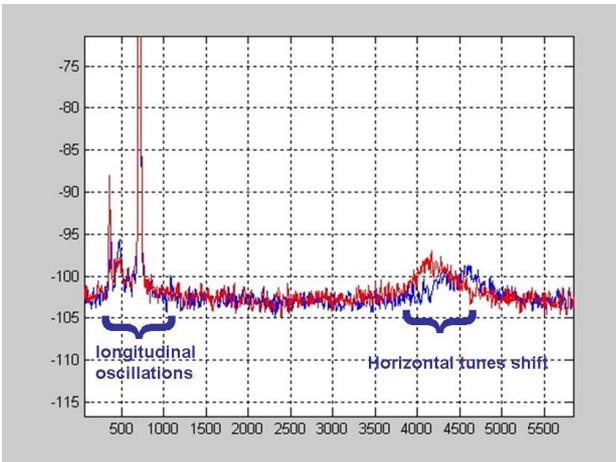

Figure 4: Zoom of the previous figure, showing the longitudinal and horizontal tunes.

Data have been recorded on September 30-th 2009; in this case the positron longitudinal feedback is turned on (as well as all the other feedback systems). The beam currents are similar to the case of November 4-th. Data coming from the spectrum analyzer are elaborated using MATLAB: the highest peak is the e+ 118-th harmonic; the red trace shows positron spectrum without collision, the blue one the same with colliding beams. The vertical scale is in dBm, the horizontal axis is in number of bins (proportional to frequency). This case is interesting because it shows a situation in which the "weak" e+ feedback has a too low gain and the "strong" electron beam damps longitudinally and shifts in frequency the positron synchrotron oscillation. In Fig. 4 it is also possible to see that the beam collisions produce also a horizontal tune shift (increasing the frequency).

With the goal to confirm the measurements done with the spectrum analyzer and to evaluate more precisely the effects, the beam diagnostic tools of the DAΦNE longitudinal feedback have been used [4]. With this system it is possible to record longitudinal data separately for each bunch. Data can be recorded both in closed and in open loop, that is with the feedback working or not.

The following Figure 5 shows the modal growth rate analysis for the cases respectively without and with collisions, turning off for a short time the bunch by bunch feedback.

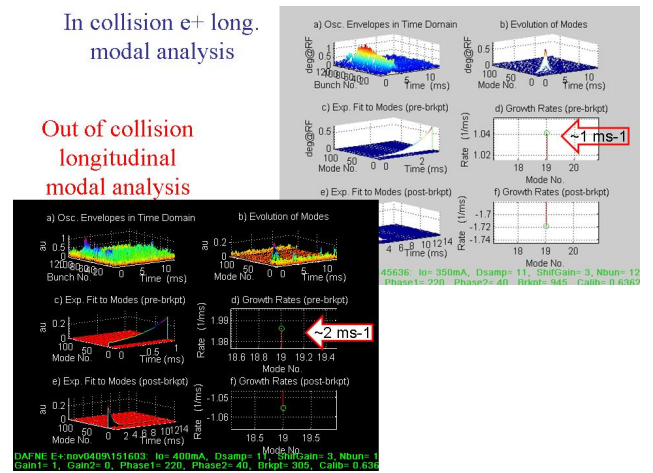

Figure 5: The Modal growth rate out of collision is 1.99 ms-1. The Modal growth rate in collision is 1.04 ms-1.

In both cases the mode 19 is the strongest unstable longitudinal mode but while out of collisions it has a growth rate, in inverse units, of 1.99 ms-1, (corresponding to 502 microseconds), in collision the growth rate is halved, 1.04 ms-1, corresponding to 961 microseconds.

Analyzing with great detail these data, it is possible to measure the synchrotron frequency shift induced on the positron beam by the beam-beam collisions with the e+ longitudinal feedback turned off.

The synchrotron frequency (out of collisions) is 34.86 kHz, while the synchrotron frequency (in collisions) is 34.23 kHz. The frequency shift induced by the beam-beam collisions is therefore -630 Hz at e+ beam currents of 320 mA (out) and 250 mA (in) for the positrons, while the e- beam currents were 520 mA (out) and 476 mA (in) respectively.

## COMMENTS

Experimental observations and measurements at DAΦNE have shown that beam-beam collisions can damp the longitudinal coupled bunch instability. Besides, a negative frequency shift of positron beam synchrotron sidebands has been observed when the beams are in collision. The authors attribute these two effects to a nonlinear longitudinal kick arising due to beam-beam interaction under a finite crossing angle. It is worthwhile to note here that we have observed this effect clearly only after implementation of the crab waist scheme of beam-beam collisions at DAΦNE having twice larger horizontal crossing angle with respect to the previous operations with the standard collision scheme [5].

In the DAΦNE technical note G-72 [6], an analytical expression for the synchrotron tune shift is outlined, and this expression gives also a measure of the synchrotron tune spread. The final formula for flat beams, that includes some additional considerations, is:

$$\xi_z = -\frac{r_e N^{strong}}{2\pi\gamma^{weak}} \frac{\left(\frac{\sigma_{z0}}{(\sigma_E/E)}\right)^{weak} tg^2(\theta/2)}{\left((\sigma_x^2 + \sigma_z^2 tg^2(\theta/2)) + \sigma_y^2\sqrt{(\sigma_x^2 + \sigma_z^2 tg^2(\theta/2))}\right)^{strong}}$$

where z is the longitudinal deviations from the synchronous "weak" particle travelling on-axis, $\sigma_x$, $\sigma_y$, $\sigma_z$ are rms sizes of the "strong" beam, the angle $\theta$ is the horizontal beam-beam crossing angle, N is the number of particles in the "strong" bunch, $r_e$ is the electron radius, $\gamma$ is the relativistic factor of the "weak" beam.

After the analytical result, the formula has been compared with numerical simulations. First of all, our simulations have confirmed that the synchrotron tune shift does not depend on parameters of the vertical motion, such as $\beta_y$ and $\nu_y$. Second, an agreement between the analytical and numerical estimates is reasonable for horizontal tunes far from integers. Quite naturally, in a scheme with a horizontal crossing angle, the synchrotron oscillations are coupled with the horizontal betatron oscillations.

One of the coupling's side effects is the $\nu_z$ dependence on $\nu_x$ that becomes stronger in vicinity of the main coupling resonances $\nu_x \pm \nu_z = k$. In order to make comparisons with the analytical formula, we need to choose the horizontal betatron tune $\nu_x$ closer to half-integer, where its influence on $\nu_z$ is weaker.

The coupling vanishes for very large Piwinski angles, and, as consequence, the $\nu_z$ dependence on $\nu_x$ is stronger for DAΦNE [7] with respect to that of SuperB [8]. Since $\nu_x$ for DAΦNE is rather close to the coupling resonance, numerical simulations have been used in order to compare the calculated synchrotron tune shift with the measured one.

In particular, when colliding, the weak positron beam with 500mA electron beam, the measured synchrotron frequency shift was about -630 Hz (peak-to-peak). In our simulations we use the DAΦNE beam parameters with respectively lower bunch current (N = 0.9 x$10^{10}$) and shorter bunch length ($\sigma_z$ = 1.6 cm). This results in the synchrotron tune shift of – 0.000232 corresponding to the frequency shift of -720 Hz. In our opinion the agreement is good considering experimental measurement errors and the finite width of the synchrotron sidebands.

## CONCLUSION

Synchrotron oscillation damping due to beam-beam collisions experimental data have been collected by a spectrum analyzer and by the bunch-by-bunch longitudinal feedback diagnostics

A simple analytical formula to explain synchrotron tune shift and tune spread due to beam-beam collisions with a crossing angle has been presented.

The formula agrees well with the simulations when the horizontal tune is far from the synchro-betatron resonances. The agreement is better for larger Piwinski angles.

Measured and obtained by simulations synchrotron frequency shifts are in a good agreement

Calculations have shown that at high beam currents the synchrotron tune spread induced by the beam-beam interaction at DAΦNE can be larger than the tune spread due to the nonlinearity of the RF voltage. This may result in additional Landau damping [9] of the longitudinal coupled bunch oscillations.